\documentclass[journal]{IEEEtran}
\usepackage{graphicx}      
\usepackage{subfig}
\usepackage[square,sort,comma,numbers]{natbib}
\usepackage{amsmath}
\usepackage{amsfonts}
\usepackage{url}
\usepackage[table]{xcolor}
\usepackage{color}
\usepackage{bm}
\let\labelindent\relax
\usepackage{enumitem}
\usepackage{verbatim}
\usepackage{subfig}

\definecolor{ao(english)}{rgb}{0.0, 0.5, 0.0}

\usepackage[utf8]{inputenc}

\begin{document}
%
\title{A simple method for shifting local \textit{dq} impedance models to a global reference frame for stability analysis}


\author{{Atle~Rygg,~Marta~Molinas,~Eneko~Unamuno,~Chen~Zhang~and~Xu~Cai}%

}

\maketitle

\begin{abstract}             
Impedance-based stability analysis in the \textit{dq}-domain is a widely applied method for power electronic dominated systems. An inconvenient property with this method is that impedance models are normally referred to their own local reference frame, and need to be recalculated when referring to a global reference frame in a given network. This letter presents a simple method for translating impedance sub-models within a complex network, from their own reference frames to any given point in the network. What distinguishes this method is that by using a simple rotational matrix, it only needs impedance models in their own local reference frames, to be translated to a global reference in the network. By way of this method, standard circuit analysis rules for series and parallel connection are applicable, as proven in the letter. The method is  defined and validated for impedances in the \textit{dq} and modified sequence domains, and it is shown that the dependency on reference frame is marginal in the latter. An additional finding from the application of this method is that components or subsystems with a certain symmetry property called \textit{Mirror Frequency Decoupling} are invariant to the choice of reference frame.

The method is illustrated and validated by comparing analytical calculations with a frequency sweep in MATLAB Simulink.

\end{abstract}


%

\section{Introduction}
Small-signal stability analysis of power electronic systems is often conducted by the impedance-based analysis \cite{middlebrook1976,Sun2009}. Previous works have developed impedance models of power electronic converters by various techniques, e.g. harmonic linearization in the sequence domain \cite{Sun2009,Cespedes2013} and \textit{dq}-domain modeling \cite{Belkhayat1997,Wen2015,Harnefors2007}. Recently, other methods proposed the modified sequence domain modeling \cite{Rygg2016,Bakhshizadeh2016,Wang2017} and phasor-based impedance modeling \cite{Shah2017}.

The \textit{dq}-domain method, modified sequence domain method and phasor-based method all share the same challenge when applied to systems with multiple units - the impedance matrices are referred to a certain reference frame or phase angle. In other words, they are referred to a certain local point in the network. Consequently, when performing system-level analysis, one must ensure that all submodels are referred to the same (global) reference frame. So far, this challenge has only been addressed by a few papers performing \textit{dq}-domain based analysis. In \cite{Burgos2016} and \cite{Cao2017}, \textit{dq}-domain stability analysis is performed to a system composed by multiple converters. The challenge of local vs. global reference frame is addressed by introducing a rotation matrix in a similar way as in the present paper, with the difference that their rotation matrix is integrated into the converter models by a case specific method, and the same method is not applicable to an arbitrarily given network. The resulting expressions for source and load subsystem impedance matrices are also very complex, and the complexity increases drastically for larger systems.

The present letter proposes a simple method in which the alignment is achieved by a rotation matrix based on load flow information. The method enables the use of impedance models in their own local reference frame without any knowledge of the internal structure. When all models are referred to the global reference frame, standard circuit analysis rules with impedance matrices can be applied. In this way, by referring all impedance matrices to a global reference frame, series and parallel connection rules are applicable. The letter also reveals that:

\begin{itemize}
  \item Sub-blocks or subsystems that satisfies the definition of a \textit{Mirror Frequency Decoupled} (MFD) system \cite{Rygg2016} are invariant to the choice of reference frame. 
  \item In the \textit{dq}-domain all four matrix elements are affected by rotation, while in the modified sequence domain, only the angles in the off-diagonal elements are affected. 
\end{itemize}

The proposed method makes \textit{dq}- and modified sequence domain impedance analysis applicable to systems of any scale. The method has low complexity and is simple to use. The letter is organized with the method presented in section \ref{sec:method}, while it is applied to a case study system in section \ref{sec:case}. The method is validated by comparing the impedance matrix from analytic calculations with a frequency sweep from a MATLAB time-domain simulation. Important mathematical derivations as well as parameter values are included in the appendix.

 \begin{figure}[h!]
  \centering
      \includegraphics[width=0.3\textwidth]{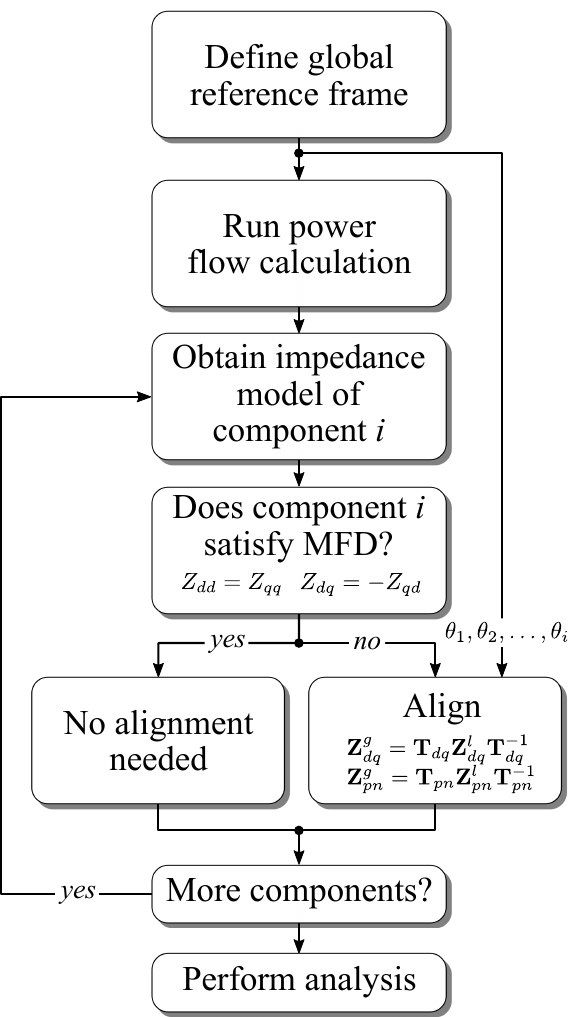}
  \caption{Flowchart to illustrate the steps to implement the proposed method}
  \label{fig:flowchart}
\end{figure}

 \begin{figure}[ht]
  \centering
      \includegraphics[width=0.4\textwidth]{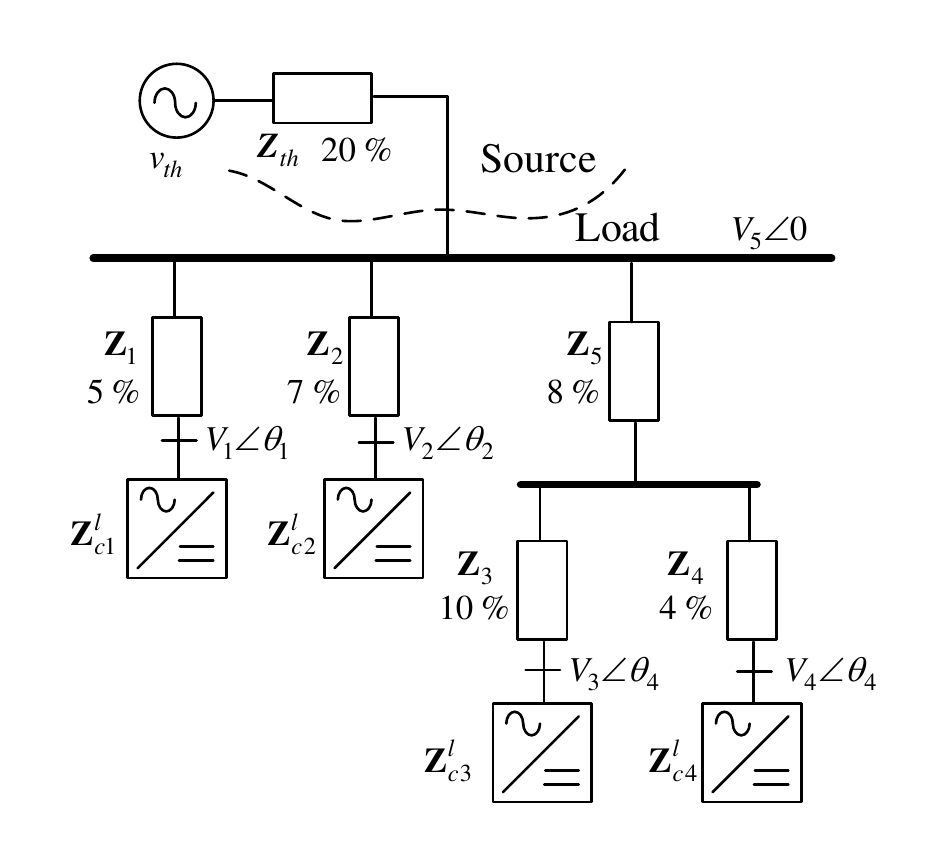}
  \caption{Overview of simulation case study}
  \label{fig:system_sketch}
\end{figure}

\section{Method description}\label{sec:method}

The method is explained based on the flowchart in Fig. \ref{fig:flowchart}, each step is explained in the following subsections.

\subsection{Define global reference frame}\label{sec:interface}
The first step in the proposed framework is to select a node which all impedance matrices shall be referred to. Any node can be selected, and the choice will not affect the stability analysis. If the selected stability analysis is based on source and load impedance equivalents, a logical choice of global reference frame is the source/load interface point.

\subsection{Run power flow calculation}
Running a power flow calculation will provide the steady-state operation point of the system. This is required for two purposes. First, analytical impedance models, e.g. the ones derived in \cite{Wen2015,Harnefors2007,Cespedes2013} are dependent on the operation point. This is due to non-linearities in the converter controller that requires linearization around the operation point. Second, the method proposed in this letter requires the information of the fundamental voltage angle at each node in the system, defined as $\theta_1,\theta_2...\theta_n$ in Fig. \ref{fig:flowchart}. This is also highlighted in the case example in Fig. \ref{fig:system_sketch}. The phase angle at the global reference frame node is defined as $\theta=0$.

\subsection{Obtain local impedance models of components}
Power electronic converters (and other units) are represented by local terminal equivalents in impedance-based analysis. The terminal equivalent is defined by the impedance model in this letter, as the analysis is based on transfer matrices between current and voltage, similar to \cite{Belkhayat1997,Wen2015,Burgos2010,Harnefors2007}. The impedance models can be obtained either from analytical models, or from measurements. Manufacturers can provide these models from e.g. factory tests.

It is important to emphasize that the models of each component are normally referred to their local terminal point, indicated by superscript \textit{l} in this letter. Referring the sub-models to a global reference frame is explained in subsection \ref{sec:align}.

Referring to the case example in Fig. \ref{fig:system_sketch}, one must obtain the impedance models of each converter $\mathbf{Z}_{c1}^l...\mathbf{Z}_{c4}^l$ and also information of other network components, e.g. impedances $\mathbf{Z}_1...\mathbf{Z}_4$ and $\mathbf{Z}_{th}$. Bold $\mathbf{Z}$ indicates that these impedances are 2x2 matrices.

\subsection{Does component satisfy MFD?}\label{sec:mfd}
A \textit{Mirror Frequency Decoupled} (MFD) system was defined in \cite{Rygg2016}, and this definition is useful when discussing local vs. global reference frames. It is shown in appendix \ref{app:rot} that MFD systems are not dependent on the reference frame, i.e. they are rotational invariant. A list of typical power system components and control blocks categorized as MFD and MFC is presented in TABLE \ref{tab:mfd}. Generally, power electronic converters are \underline{not} MFD due to control blocks such as PLL and DC-link voltage controllers. Passive components are always MFD.

\begin{table}[h!]
    \centering
    \caption{MFD categorization of typical components and control system blocks}
    \begin{tabular}{c|c}
         \textbf{Mirror Frequency Decoupled} & \textbf{Mirror Frequency Coupled } \\ \hline
         Linear passive elements (R,L,C)   & Phase Lock Loop (PLL) \\
         \textit{dq} current controller      & DC-link voltage controller \\
         $\alpha \beta$ controllers          & Salient-pole machines \\
         Transformers and cables             & Active and reactive power controllers \\
         Round-rotor machines
    \end{tabular}
        \label{tab:mfd}
\end{table}

\subsection{Align models with global reference frame}\label{sec:align}
If a component does \textit{not} satisfy the MFD condition, its impedance model will depend on the reference frame. The needed alignment from local to global reference frame is given by the following relations derived in appendix \ref{app:rot}: 
\begin{align}
    &\mathbf{Z}_{dq}^{g} = \mathbf{T}_{dq} \mathbf{Z}_{dq}^{l} \mathbf{T}_{dq}^{-1} \nonumber \\
    &\mathbf{Z}_{pn}^{g} = \mathbf{T}_{pn} \mathbf{Z}_{pn}^{l} \mathbf{T}_{pn}^{-1} \nonumber \\
    \mathbf{T}_{dq}&= \begin{bmatrix} \cos \theta_i & \sin \theta_i \\-\sin \theta_i & \cos \theta_i \end{bmatrix}
     \qquad
    \mathbf{T}_{pn}= \begin{bmatrix} e^{j\theta_i} & 0 \\0 & e^{-j\theta_i} \end{bmatrix}
    \label{eq:rot_dq_pn}
\end{align}
where subscript \textit{dq} denotes a \textit{dq}-domain model, while \textit{pn} denotes a modified sequence domain model. Superscript \textit{g} denotes global reference frame, while \textit{l} denotes local reference frame. The angle $\theta_i$ is the angle between component \textit{i} local reference frame and the global reference frame defined in section \ref{sec:interface}.

Of note, all four matrix elements are affected by rotation in the \textit{dq}-domain, while only the angles of the off-diagonal elements are affected by rotation in the modified sequence domain.

A special case is when there is no power flow in the system (no load case). Then, all voltage angles are equal, and no rotation is needed.

\subsection{Perform analysis}

Once impedance models are obtained for all network components in the system, and referred to the global reference frame, the stability analysis can be conducted by various methods. A common approach is to apply the Generalized Nyquist Criterion (GNC) to the source and load impedance equivalents. This was first applied to power electronic systems in \cite{Belkhayat1997}. See \cite{Wen2015} and \cite{Cao2017} for examples of GNC applied to power electronic systems. 

When applying GNC or other analysis methods, it is important that all submodels are aligned with the same reference frame. As an example, (\ref{eq:Z_load}) presents how a equivalent impedance of a four converter subsystem is calculated by series and parallel connection.

\begin{figure}[!ht]
\minipage{0.48\textwidth}
  \includegraphics[width=1\textwidth]{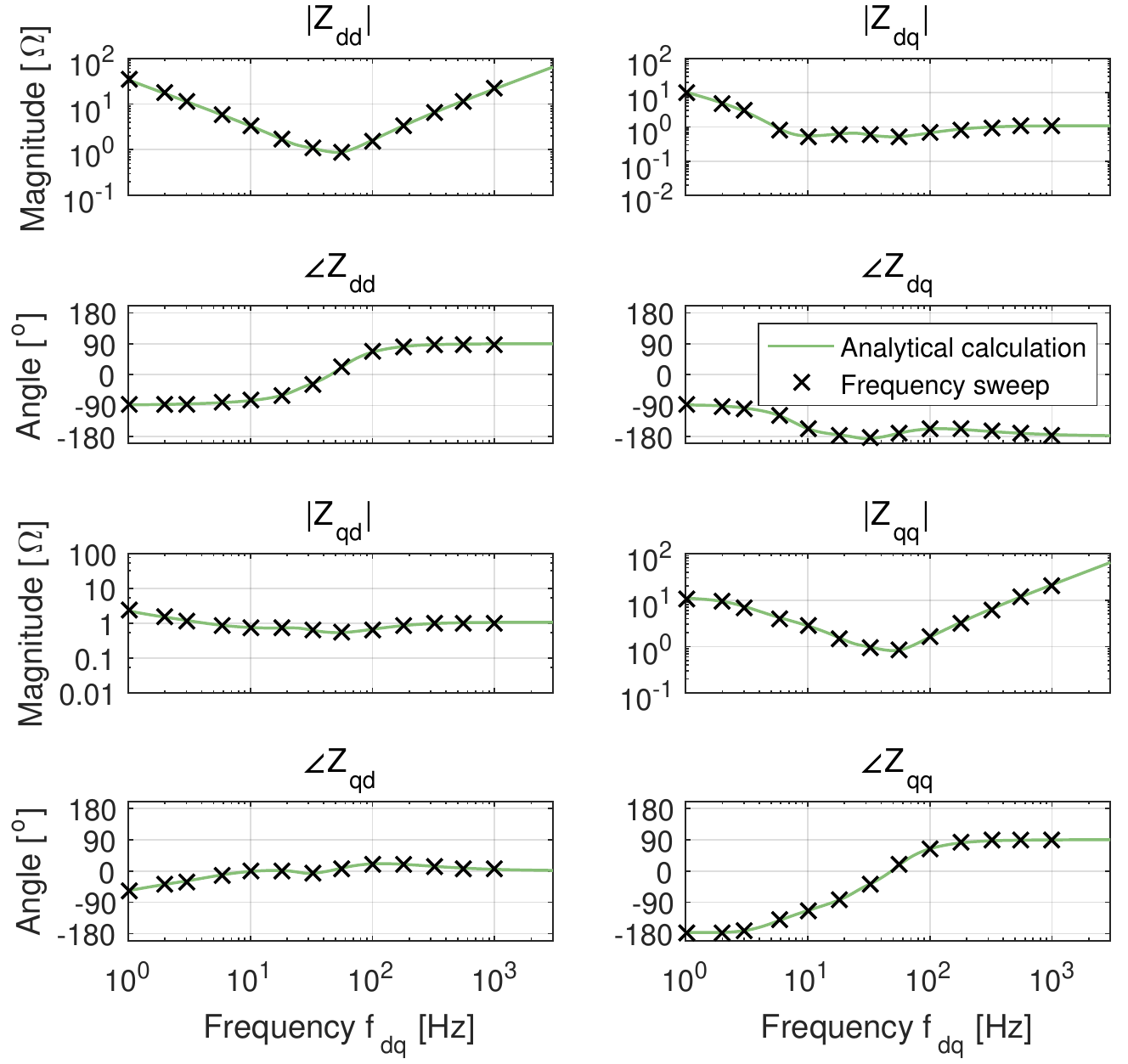}
  \caption{Comparison of analytical model (\ref{eq:Z_load}) with frequency sweep: \textit{dq}-domain}
  \label{fig:result_dq}
\endminipage\hfill
\minipage{0.48\textwidth}
      \includegraphics[width=1\textwidth]{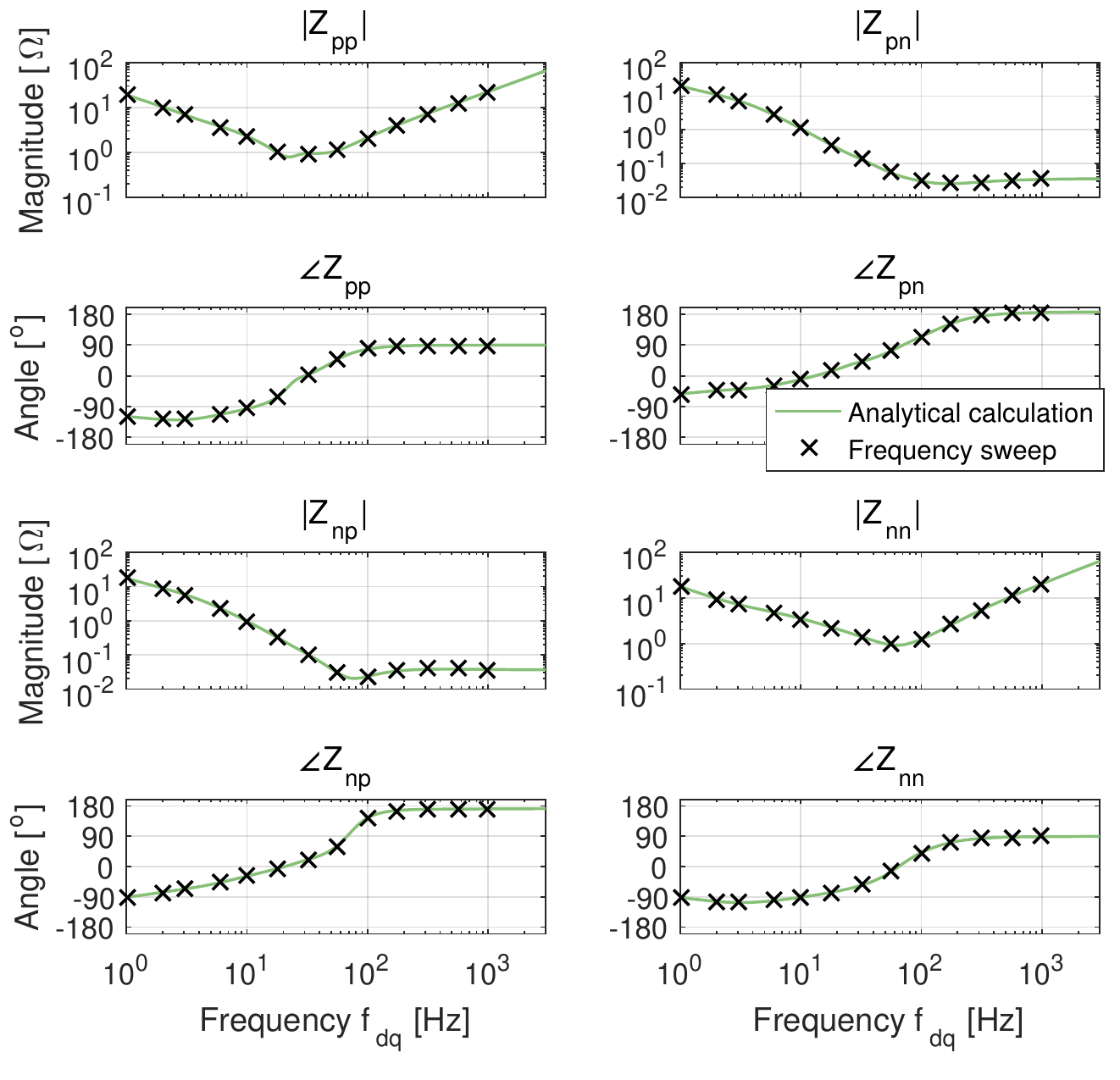}
  \caption{Comparison of analytical model (\ref{eq:Z_load}) with frequency sweep: \textit{pn}-domain}
  \label{fig:result_pn}
\endminipage\hfill
\end{figure}

\section{Validation by simulation}\label{sec:case}
A network example is presented in Fig. \ref{fig:system_sketch}. The system includes four grid-connected converters connected in a radial structure. Details on the converter model is given in Appendix \ref{app:params}. The analytical model provides the impedance matrices $\mathbf{Z}_{c1}^l...\mathbf{Z}_{c4}^l$ shown in Fig. \ref{fig:system_sketch}.

The load subsystem total impedance can be found by series and parallel connection as: 
\begin{equation}
    \mathbf{Z}_{load} = (\mathbf{Z}_1+\mathbf{Z}_{c1}^g)||(\mathbf{Z}_2+\mathbf{Z}_{c2}^g)
    ||(\mathbf{Z}_5+(\mathbf{Z}_3+\mathbf{Z}_{c3}^g)||(\mathbf{Z}_4+\mathbf{Z}_{c4}^g]))
    \label{eq:Z_load}
\end{equation}
where all converter impedances are referred to the global reference frame by applying the relations in (\ref{eq:rot_dq_pn}) to the local impedance models. This is indicated by superscript \textit{g}. Note that the line impedances do not require alignment with global reference frame as they satisfy the MFD condition.

The comparison between the analytical calculation from (\ref{eq:Z_load}) and a frequency sweep in MATLAB simulink is presented in Fig. \ref{fig:result_dq} for the \textit{dq}-domain and Fig. \ref{fig:result_pn} for the modified sequence domain. The perturbation injection method is applied at the interface point in Fig. \ref{fig:system_sketch}. It is clear that the impedances resulting from the analytic calculation have an exact match with the ones resulting from the frequency sweep,  validating the relations in (\ref{eq:rot_dq_pn}) and by that the method for referring impedance matrices from local to a global reference frame.

\section{Conclusion}\label{sec:conclusion}
The letter has presented a simple method to ease the applicability of impedance-based analysis in the \textit{dq}-domain and modified sequence domain, at different points in a given complex network. By a simple rotation matrix, the different impedance models in the network can be moved from their local reference frames to a global reference frame within the network. The decoupled nature of the method prevents the need to integrate the rotation factor into the local impedance model and by that greatly simplifies the system analysis referred to a global reference frame. A secondary result of this method is that by way of the proposed rotation matrix, aggregate impedances can be easily calculated based on impedance sub-models referred to their own local terminals. The method then allows the use of standard series and parallel connection rules, which makes system analysis significantly easier.

It is found that in the \textit{dq}-domain all four impedance elements are affected by rotation, while in the modified sequence domain only the off-diagonal elements are affected. Furthermore, it is shown that a \textit{Mirror Frequency Decoupled} (MFD) system is invariant to rotation, and can therefore be used directly in series and parallel connection.

\small
\bibliographystyle{IEEEtran}
\bibliography{PhD-arbeid}

\appendix

\subsection{\textit{dq} symmetric systems and mirror frequency decoupled systems}\label{app:mfd}

In \cite{Rygg2016} a system was defined as \textit{Mirror Frequency Decoupled} (MFD) if there is no coupling between $s+j\omega_1$ and $s-j\omega_1$, that is $Z_{pn}=Z_{np}=0$. It can be shown that this property is equal to a \textit{dq symmetric} system as defined in \cite{Harnefors2007b}. This condition is given by:
\begin{equation}
    Z_{dd} = Z_{qq} \qquad Z_{dq}=-Z_{qd}
    \label{eq:dqsymm}
\end{equation}

It was shown in \cite{Rygg2016} that in a MFD system the original sequence domain impedance defined in \cite{Sun2009} will give identical stability analysis as the \textit{dq}-domain analysis. 

\subsection{Derivation of rotation matrices}\label{app:rot}
Rotating a \textit{dq}-domain voltage (or current) vector with an angle $\theta$ is achieved by the following relation \cite{Cao2017}:
\begin{equation}
    \begin{bmatrix}V_d \\ V_q\end{bmatrix}_{rot} = 
    \begin{bmatrix} \cos \theta & -\sin \theta \\ \sin \theta & \cos \theta \end{bmatrix}
    \begin{bmatrix}V_d \\ V_q\end{bmatrix}
    =\mathbf{T}_{dq}\begin{bmatrix}V_d \\ V_q\end{bmatrix}
\end{equation}
where the rotation matrix is defined as $\mathbf{T}_{dq}$. As the same equation is valid for the current vector, the \textit{dq}-domain impedance matrix is rotated as:
\begin{equation}
    \mathbf{Z}_{dq,rot}=\mathbf{T}_{dq}\mathbf{Z}_{dq}\mathbf{T}_{dq}^{-1}
\end{equation}

By applying the impedance transform from \cite{Rygg2016}, the rotation matrix $\mathbf{T}_{pn}$ in the modified sequence domain is found as:
\begin{align}
    \mathbf{T}_{pn}&=\mathbf{A}_z \mathbf{T}_{dq} \mathbf{A}_z^{-1} = \frac{1}{2} \begin{bmatrix}1 & j \\1 & -j\end{bmatrix}
        \begin{bmatrix} \cos \theta & -\sin \theta \\ \sin \theta & \cos \theta \end{bmatrix}
        \begin{bmatrix}1 & 1 \\-j & j\end{bmatrix} \nonumber \\
    &=\begin{bmatrix}e^{j \theta} & 0 \\0&  e^{-j \theta} \end{bmatrix}
\end{align}

Expanding this rotation matrix yields:
\begin{equation}
    \mathbf{Z}_{pn,rot} = \mathbf{T}_{pn}\mathbf{Z}_{pn}\mathbf{T}_{pn}^{-1}
    =\begin{bmatrix}Z_{pp} & Z_{pn}e^{j2\theta} \\ Z_{np}e^{-j2\theta} & Z_{nn} \end{bmatrix}\label{eq:Zpn_rot}
\end{equation}

It is seen from (\ref{eq:Zpn_rot}) that only the angle in the off-diagonal elements $Z_{pn}$ and $Z_{np}$ are affected by rotation in the modified sequence domain. As these elements are zero for a MFD-system (appendix \ref{app:mfd}), a MFD system is rotational invariant in the modified sequence domain. As a consequence, the \textit{dq}-domain impedance matrix is also invariant to rotation for a MFD system.

 \begin{figure}[ht!]
  \centering
      \includegraphics[width=0.4\textwidth]{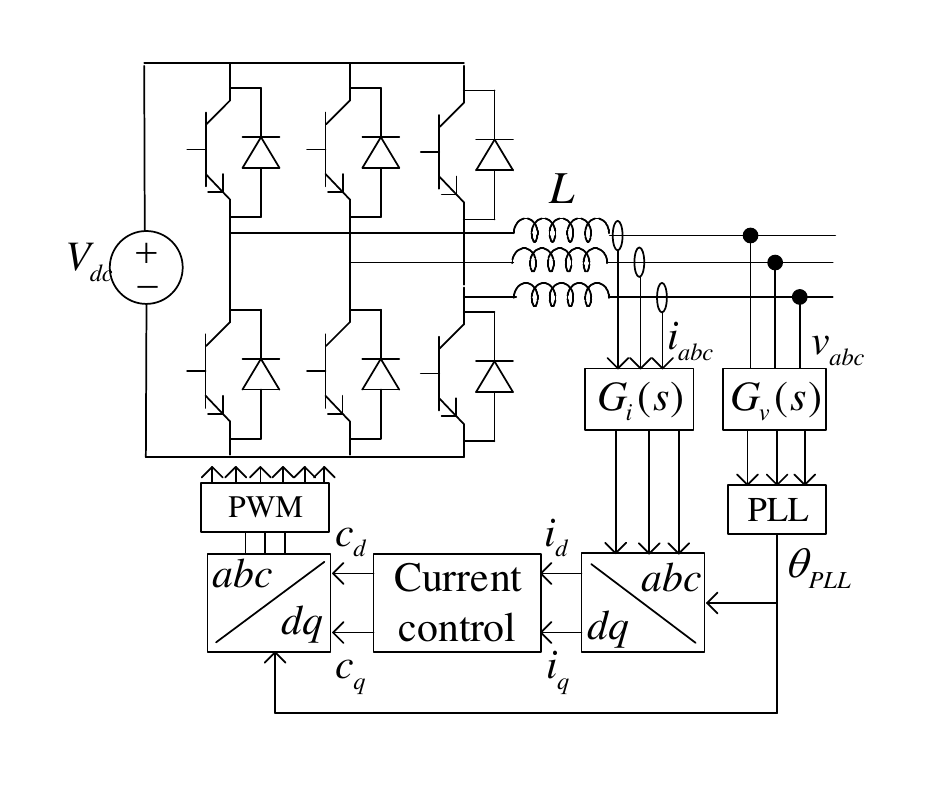}
  \caption{Overview of converter model}
  \label{fig:converter_sketch}
\end{figure}

\begin{table}[h!]
    \centering
    \begin{tabular}{l | l}
        $V_{th}$=  $6600$ $V$ LL-RMS   &   $S_{base}$ = $10$ $MW$\\
        $f_{n}$=  $50$ $Hz$ & $Z_{base}$ = $4.36$ $\Omega$ \\
        $k_{p}$ = $6.43\cdot 10^{-4}$ $p.u./A$  & $k_{i}$=  $0.161$ $p.u./(As)$   \\
        $k_{p,pll}$ = $0.00758$ $rad/(Vs)$  &  $k_{i,pll}$ = $0.152$ $rad/(Vs^2)$  \\
        $V_{dc}$=  $1127$ $V$ & $G_i(s)=G_v(s)=\frac{1}{1+5\cdot 10^{-4}s}$   \\
        $L_{conv}=$ = 6.93 $mH$   &  
    \end{tabular}
    \caption{Parameter values applied in the simulation case study}
    \label{tab:setpoints}
\end{table}

\begin{table}[h!]
    \centering
    \begin{tabular}{l | l | l |l |l |l } 
        &       $\mathbf{Z}_{c1}$ & $\mathbf{Z}_{c2}$ & $\mathbf{Z}_{c3}$ & $\mathbf{Z}_{c4}$  \\ \hline
        $I_d$ & 100 A & 130 A & 80 A & 150 A \\
        $I_q$ & 33 A & 43 A & 39 A & 0 A \\     
        $\theta$    & $0.97^o$ & $2.53^o$ & $4.15^o$ & $4.89^o$ 
    \end{tabular}
    \caption{Operation points for the converters, including their terminal voltage steady-state fundamental angles $\theta$}
    \label{tab:params}
\end{table}

\subsection{Case study data}\label{app:params}

The converter model used in the case study is presented in Fig. \ref{fig:converter_sketch}. The current controller is a standard PI-controller $H_c(s) = k_p + \frac{k_i}{s}$, while the PLL is a standard synchronous reference frame PLL based on a PI-controller $H_{PLL}(s)=k_{p,pll}+\frac{k_{i,pll}}{s}$. Numerical data is given in table \ref{tab:params}. The set-points for each current controller are given in Table \ref{tab:setpoints}. The impedance model for each converter is based on the model derived in \cite{Wen2015}, and repeated in (\ref{eq:wen}). Expressions for each transfer matrix in (\ref{eq:wen}) are omitted here but can be found in \cite{Wen2015}. The corresponding modified sequence domain model is obtained by the transform derived in \cite{Rygg2016}.
\begin{align}
   \mathbf{Z}_{c,dq}=&(Z_{out}^{-1}+G_{id}G_{del}([-G_{ci}+G_{dei}]G_{PLL}^i + G_{PLL}^d]K)^{-1}\cdot \nonumber \\
   &(I+G_{id}G_{del}[G_{ci} -G_{dei}] K)
   \label{eq:wen}
\end{align}
The line impedances are given as per unit values in Fig. \ref{fig:system_sketch}, and the per unit base $Z_{base}$ is given in Table \ref{tab:params}. The X/R-ratio is 10 for each impedance.

\end{document}